\documentclass[onecolumn,ltaffilletter,superscriptaddress,notitlepage]{revtex4-1}
\usepackage{graphicx,url,amssymb,amsmath,rotating,color,units,wasysym,epsfig,multirow,epstopdf,subcaption,xcolor}
\usepackage[colorlinks,urlcolor=blue,citecolor=blue,linkcolor=blue]{hyperref}
\usepackage{soul,xcolor}

\begin{document}

%\title{}
% remove page number from title page
\thispagestyle{empty}

\begin{figure*}
    \centering
    \includegraphics[width=1.0\columnwidth]{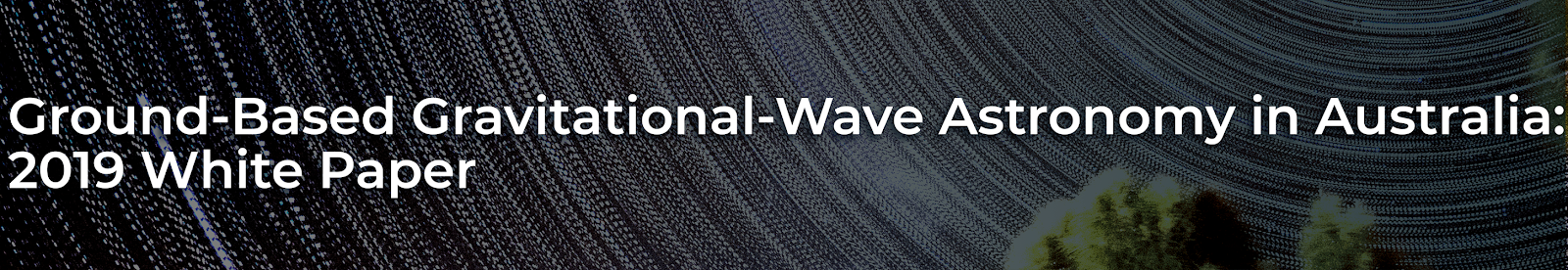}
\end{figure*}

\author{Matthew Bailes}
\affiliation{Centre for Astrophysics and Supercomputing, Swinburne University of Technology, Hawthorn, VIC 3122, Australia}
\affiliation{OzGrav: The ARC Centre of Excellence for Gravitational-Wave Discovery}

\author{David McClelland}
\affiliation{Department of Quantum Science, Research School of Physics, The Australian National University, Acton, ACT 2601, Australia}
\affiliation{OzGrav: The ARC Centre of Excellence for Gravitational-Wave Discovery}

\author{Eric Thrane}
\email{eric.thrane@monash.edu}
\affiliation{School of Physics and Astronomy, Monash University, Clayton, VIC 3800, Australia}
\affiliation{OzGrav: The ARC Centre of Excellence for Gravitational-Wave Discovery}

\author{David Blair}
\affiliation{Department of Physics, The University of Western Australia, Crawley, WA 6009, Australia}
\affiliation{OzGrav: The ARC Centre of Excellence for Gravitational-Wave Discovery}

\author{Jeffrey Cooke}
\affiliation{Centre for Astrophysics and Supercomputing, Swinburne University of Technology, Hawthorn, VIC 3122, Australia}
\affiliation{OzGrav: The ARC Centre of Excellence for Gravitational-Wave Discovery}

\author{David Coward}
\affiliation{Department of Physics, The University of Western Australia, Crawley, WA 6009, Australia}
\affiliation{OzGrav: The ARC Centre of Excellence for Gravitational-Wave Discovery}

\author{Robin Evans}
\affiliation{School of Electrical and Electronic Engineering, The University of Melbourne, Parkville, VIC 3010, Australia}
\affiliation{OzGrav: The ARC Centre of Excellence for Gravitational-Wave Discovery}

\author{Yeshe Fenner}
\affiliation{Centre for Astrophysics and Supercomputing, Swinburne University of Technology, Hawthorn, VIC 3122, Australia}
\affiliation{OzGrav: The ARC Centre of Excellence for Gravitational-Wave Discovery}

\author{Duncan Galloway}
\affiliation{School of Physics and Astronomy, Monash University, Clayton, VIC 3800, Australia}
\affiliation{OzGrav: The ARC Centre of Excellence for Gravitational-Wave Discovery}

\author{Jarrod Hurley}
\affiliation{Centre for Astrophysics and Supercomputing, Swinburne University of Technology, Hawthorn, VIC 3122, Australia}
\affiliation{OzGrav: The ARC Centre of Excellence for Gravitational-Wave Discovery}

\author{Li Ju}
\affiliation{Department of Physics, The University of Western Australia, Crawley, WA 6009, Australia}
\affiliation{OzGrav: The ARC Centre of Excellence for Gravitational-Wave Discovery}

\author{Paul Lasky}
\affiliation{School of Physics and Astronomy, Monash University, Clayton, VIC 3800, Australia}
\affiliation{OzGrav: The ARC Centre of Excellence for Gravitational-Wave Discovery}

\author{Ilya Mandel}
\affiliation{School of Physics and Astronomy, Monash University, Clayton, VIC 3800, Australia}
\affiliation{OzGrav: The ARC Centre of Excellence for Gravitational-Wave Discovery}

\author{Kirk McKenzie}
\affiliation{Department of Quantum Science, Research School of Physics, The Australian National University, Acton, ACT 2601, Australia}
\affiliation{OzGrav: The ARC Centre of Excellence for Gravitational-Wave Discovery}

\author{Andrew Melatos}
\affiliation{School of Physics, The University of Melbourne, Parkville, VIC 3010, Australia}
\affiliation{OzGrav: The ARC Centre of Excellence for Gravitational-Wave Discovery}

\author{David Ottaway}
\affiliation{Department of Physics and The Institute of Photonics and Advanced Sensing (IPAS), The University of Adelaide, Adelaide, SA, 5005, Australia}
\affiliation{OzGrav: The ARC Centre of Excellence for Gravitational-Wave Discovery}

\author{Susan Scott}
\affiliation{Department of Quantum Science, Research School of Physics, The Australian National University, Acton, ACT 2601, Australia}
\affiliation{OzGrav: The ARC Centre of Excellence for Gravitational-Wave Discovery}

\author{Bram Slagmolen}
\affiliation{Department of Quantum Science, Research School of Physics, The Australian National University, Acton, ACT 2601, Australia}
\affiliation{OzGrav: The ARC Centre of Excellence for Gravitational-Wave Discovery}

\author{Peter Veitch}
\affiliation{Department of Physics and The Institute of Photonics and Advanced Sensing (IPAS), The University of Adelaide, Adelaide, SA, 5005, Australia}
\affiliation{OzGrav: The ARC Centre of Excellence for Gravitational-Wave Discovery}

\author{Linqing Wen}
\affiliation{Department of Physics, The University of Western Australia, Crawley, WA 6009, Australia}
\affiliation{OzGrav: The ARC Centre of Excellence for Gravitational-Wave Discovery}

\author{Chunnong Zhao}
\affiliation{Department of Physics, The University of Western Australia, Crawley, WA 6009, Australia}
\affiliation{OzGrav: The ARC Centre of Excellence for Gravitational-Wave Discovery}

%\date{November 2019}
%\begin{abstract}
%\end{abstract}

\maketitle

\section*{Executive Summary}
The past four years have seen a scientific revolution through the birth of a new field: gravitational-wave astronomy. The first detection of gravitational waves~\cite{gw150914}---recognised by the 2017 Nobel Prize in Physics---provided unprecedented tests of general relativity while unveiling a previously unknown class of massive black holes, thirty times more massive than the Sun. The subsequent detection of gravitational waves from a merging binary neutron star~\cite{GW170817} confirmed the hypothesised connection between binary neutron stars and short gamma-ray bursts~\cite{GRB170817A} while providing an independent measurement of the expansion of the Universe~\cite{Hubble}. The discovery enabled precision measurement of the speed of gravity~\cite{GRB170817A} while shedding light on the origin of heavy elements~\cite{GW170817_kilonova}. At the time of writing, the Laser Interferometer Gravitational-wave Observatory (LIGO)~\cite{LIGO} and its European partner, Virgo~\cite{Virgo}, have published the detection of eleven gravitational-wave events~\cite{gwtc-1}. New, not-yet-published detections are announced on a nearly weekly basis. This fast-growing catalogue of gravitational-wave transients is expected to yield insights into a number of topics, from the equation of state of matter at supra-nuclear densities~\cite{GW170817_eos} to the fate of massive stars~\cite{o2_pop}.

With the field of gravitational-wave astronomy now firmly established, the international community is beginning to chart out a course for the coming decades. In the short term, \$16M USD in funding has been awarded to upgrade Advanced LIGO to a new facility called A+, set to begin taking data in 2023~\cite{Reitze}. Through a series of ARC LIEF grants, Australia is contributing key A+ components and commissioning staff. The A+ observatory will detect binary neutron stars more than three times as far away as Advanced LIGO (i.e. $\sim$30$\times$ as often), to redshifts of z = $\sim$0.4~\cite{Reitze}. Nearby binary neutron stars will be detected in gravitational waves up to ten minutes before they merge~\cite{Reitze}. In the longer term, the international community is moving to begin engineering studies for third-generation (3G) observatories such as the Einstein Telescope and Cosmic Explorer.

These facilities---which could begin taking data in the mid 2030s---will enable transformational science. Cosmic Explorer will detect binary neutron stars out to redshifts of z=12 when the Universe was just $\unit[400]{Myr}$ old~\cite{Reitze}. (This is beyond the most distant known object, a galaxy at z=11.09~\cite{Oesch}.) Cosmic Explorer could also see binary black holes like GW150914 out to z=37~\cite{Reitze}---well before stellar-mass binary black holes are thought to have even formed. Third-generation observatories like Cosmic Explorer will complement existing Australian investments. In particular, the Square Kilometre Array (SKA)~\cite{Dewdney2016,Braun2017} will enable multi-messenger analysis of gravitational-wave sources identified by LIGO and the Laser Interferometer Space Antenna (LISA)~\cite{LISA} while probing nanohertz gravitational waves with pulsar timing.

The science potential of 3G observatories is enormous, enabling measurements of gravitational waves from the edge of the Universe and precise determination of the neutron star equation of state. Australia is well-positioned to help develop the required technology. {\bf The mid-term review should consider investment in a scoping study for an Australian Gravitational-Wave Pathfinder that develops and validates core technologies required for the global 3G detector network.} This pathfinder will put Australia in a position to host a major facility while creating the technology necessary for the global 3G network. Estimated cost through 2025: \$5M AUD.

{\bf The mid-term review should consider prioritising support for the Gravitational Wave Data Centre at or above the current funding levels through 2026 to establish an Australian LIGO Tier-2 data centre ready for A+.} This support will enable Australia to consolidate the increased influence gained since the detection of gravitational waves, leading the science enabled by Advanced LIGO and A+, while growing capacity for gravitational-wave analysis of data with radio telescopes such as the SKA. Estimated cost through 2025: \$4.4M AUD.

\section*{Science Overview}
We advocate investment in an Australian Gravitational-Wave Pathfinder that develops and validates core technologies required for the global 3G detector network. The international community has identified six key 3G science goals, that we summarize here~\cite{Reitze}:
\begin{enumerate}
    \item {\bf Extreme matter.} Neutron stars are made of the densest matter in the Universe. The behaviour of matter at neutron-star densities is an unsolved problem in nuclear physics. As two neutron stars merge, they bulge and buckle under the influence of each other's gravity, leaving a tell-tale signature in the gravitational waveform. Measurements of gravitational waves from merging neutron stars will provide insights into the nature of matter at supra-nuclear densities. (Decadal plan research question \#6.)
    \item {\bf Multi-messenger astronomy.} The multi-messenger detection of the merging binary neutron star GW170817 provides an inkling of the science possible with 3G observatories. The event GW170817 is probably the most-studied astronomical event in history, allowing us to study the nature of gamma-ray bursts, kilonovae, shock physics, the Hubble expansion, etc. The same event observed in a 3G network could be detected 90 minutes~\cite{Reitze} before merger with dramatically improved sky localisation, enabling a wealth of electromagnetic observations not possible with GW170817. (Decadal plan research questions \#2,3.)
    \item {\bf The high-redshift Universe.} By probing merging binaries out to redshifts of z=37~\cite{Reitze}, 3G observatories will track the evolution of the Universe over cosmic time free from selection bias. These measurements will provide crucial input for the mission laid out by the current Astro3D Centre of Excellence: to study ``the evolution of the matter, elements and light in the Universe from the Big Bang to the present day.'' (Decadal plan research questions \#2,3.)
    \item {\bf Extreme gravity.} Cosmic Explorer will detect binary mergers with signal-to-noise ratios of thousands (up to a hundred times more than Advanced LIGO)~\cite{Reitze}. These precision measurements of gravitational waves will enable us to measure subtle general relativistic effects while looking for signs of new physics. (Decadal plan research question \#6.)
    \item {\bf Massive stars.} In one year, Cosmic Explorer is likely to detect on the order of a million binary neutron star mergers. We will measure the mass and spin of essentially every merging stellar-mass black hole in the Universe. This wealth of data will enable breakthroughs in our understanding of how stars are born and die. (Decadal plan research question \#4.)
    \item {\bf The Unexpected.} History shows that opening a new window on the Universe can reveal unexpected treasures. The serendipitous detection of gamma-ray bursts and fast radio bursts are cases in point. By building a gravitational-wave observatory with unparalleled sensitivity, we may observe something unexpected and important.
\end{enumerate}

The science case for increased support for the Gravitational Wave Data Centre is built on these same six goals: extreme matter, multi-messenger astronomy, etc. However, with the Data Centre, we can begin pursuing these goals using data from Advanced LIGO and A+. The breakthroughs enabled by Advanced LIGO's first two observing runs have established the tremendous potential of existing ``second-generation'' observatories. Continued improvements in sensitivity, along with Australian-led breakthroughs in computing and analysis, will spawn the next generation of discoveries.

\section*{Synergies}
\begin{enumerate}
\item {\bf Radio Astronomy.} Australia has made a large investment in ASKAP~\cite{askap}, the MWA~\cite{mwa}, and the SKA. We have long been a pioneer of radio astronomy techniques and facilities. The SKA will have the sensitivity to detect most of the kilonovae created from nearby merging binary neutron stars. These observations will help to uncover the origin of heavy elements, which are thought to be created in kilonovae, while precisely determining the expansion of the Universe. Moreover, facilities such as Parkes~\cite{ppta}, MeerKat, and the SKA probe low-frequency gravitational waves that are inaccessible with LIGO/Virgo. While LIGO/Virgo probe stellar-mass black holes, pulsar timing arrays create Galaxy-sized detectors with the potential to measure nanohertz gravitational waves from supermassive black holes. Australia is a pioneer of millisecond pulsar discovery and pulsar timing arrays.

\item {\bf LISA.} In 2034 the Laser Interferometer Space Antenna (LISA) observatory is scheduled to launch. LISA will revolutionise gravitational-wave astronomy by measuring millihertz gravitational waves in the band between the ground-based facilities like Cosmic Explorer and the nanohertz band of pulsar timing arrays like the SKA. Australia boasts a number of members of the LISA Consortium. Although the contracts for building LISA have been awarded, many of the techniques for the data analysis are currently being pioneered using pulsar timing and ground-based detectors. LISA will enable multi-wavelength study of the gravitational-wave sky.

\item {\bf ESO and ELTs.} It has long been a national priority to gain full membership in the European Southern Observatory (ESO), and to provide access to Australian astronomers to ~30 meter-class telescopes such as the GMT or the ELT. Gravitational-wave astronomy has one astounding advantage over electromagnetic astronomy in that linear improvements in strain sensitivity provide linear increases in range. As a result, there will soon be demand for 8 meter and greater apertures to follow up the host galaxies of gravitational-wave detections.

\item {\bf ASTRO3D, OzGrav, and the Centre of Excellence for Dark Matter Particle Physics.} These three ARC Centres of Excellence share common goals. Using different means, they seek to study how the Universe evolves, and how gravity works. These shared goals make the Centres natural collaborators.
\end{enumerate}

\section*{Engagement model}
{\bf Part One: Cosmic Explorer South.} The Gravitational Wave International Committee (GWIC) has identified the need for three third-generation observatories: the Einstein Telescope in Europe, Cosmic Explorer in the USA, and one observatory in Australasia; see Figure 1. Given the abundance of suitable surface sites, Australia's close relationship with the USA, and the economy of scale associated with common infrastructure design and development, we propose to contribute to the design and testing of 3G observatories in close partnership with our existing collaborators via Cosmic Explorer. There are two facets of this collaboration. First, we endeavour to design and test key components of Cosmic Explorer technology. Second, we aim to undertake an engineering study for a southern-hemisphere observatory, Cosmic Explorer South. Cosmic Explorer South would be funded by an international consortium, led by Australia, and include Asian countries, with potential contributions from the USA, UK and Europe.

\begin{figure*}
    \centering
    \includegraphics[width=1.0\columnwidth]{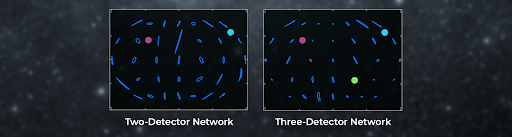}
    \caption{Gravitational-wave sky maps (adapted from~\cite{zhao-wen}). The left-hand side shows the localisation provided by a two-detector Cosmic Explorer network (for a binary neutron star merger located at a redshift of z=2) while the right-hand shows the improved localisation made possible with a three-detector network. Each blue ellipse is a localisation region for a different binary merger. The filled-in circles represent the locations of gravitational-wave observatories in a global network: Europe (purple), USA (cyan), and Australia (green). Each year, the three-detector network will localise over a thousand binary neutron stars to better than one square degree.}
\end{figure*}

In Figure 2, we show the timeline and indicative funding profile over three phases as proposed by the Cosmic Explorer team. There are two stages: Stage 1 is a room-temperature observatory, which employs the same technology planned for A+, except with a ten-times longer interferometer. Stage 2 uses new technologies, pushing for a further factor of ten improvement in strain sensitivity. It is this technological leap, which would enable Cosmic Explorer to observe every stellar-mass binary black hole in the Universe.

\begin{figure*}
    \centering
    \includegraphics[width=1.0\columnwidth]{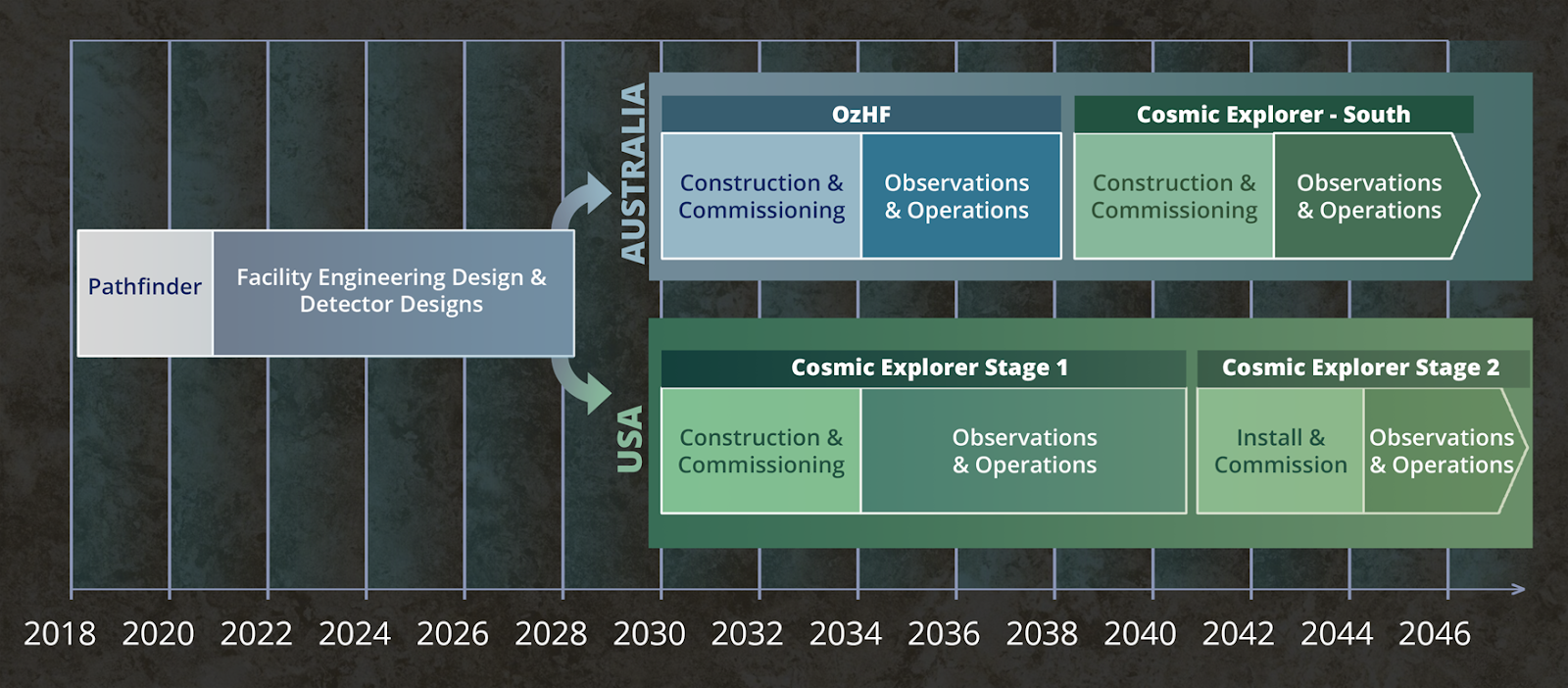}
    \caption{Timeline for R\&D and detector construction.}
\end{figure*}

The Australian timeline flips the order of operations. We propose to demonstrate Stage 2 technology on a smaller scale, developing crucial technology and infrastructure for Cosmic Explorer while enabling science goal \#1 (Extreme Matter) much sooner than otherwise possible. The first step in this program is the Australian 3G Gravitational-Wave Pathfinder:
\begin{itemize}
    \item {\bf Stage 0 (present--2029).} Participate in the international team working on the design and costing for Cosmic Explorer. The US Decadal Astronomy Plan budgets \$60M USD for this activity. We estimate that Australia can take part in the design and costing of Cosmic Explorer at a cost of \$1M AUD/yr for five years, followed by \$1.5M AUD/yr over the subsequent five years---about 10\% of the US contribution. As part of this research and development, we aim to design and cost a new facility: OzHF (see Figure 3). OzHF is a concept for a short (~4 km), high-frequency observatory in regional Australia, which would perform novel science on the unexplored state of nuclear matter at extreme densities, far beyond the sensitivity of current observatories. It would also validate key technologies for 3G detectors, helping to de-risk those facilities. At this time, the cost of OzHF is highly uncertain, but likely to be between \$100M-\$200M. At the end of this study, the feasibility of OzHF will have been established, and the cost and annual operating budget will be known to within a 10\% contingency.
    \item {\bf Stage 1 (2030--2037).} Pending successful outcome of the Stage 0, begin construction of OzHF in parallel to US construction of Cosmic Explorer North Stage 1. While OzHF is expected to be more than ten times smaller than Cosmic Explorer, with a substantially smaller budget, preliminary designs suggest that it may be able to achieve comparable sensitivity at high gravitational-wave frequencies (2-4 kHz) where the effects of Extreme Matter (science goal \#1) become important; see Figure 4. In this way, OzHF presents an opportunity to leapfrog Stage 1 science while developing infrastructure needed for Stage 2. Key specifications for OzHF and Cosmic Explorer are summarised in Figure 5.
    \item {\bf Stage 2 (2038--2046).} Pending significant investment from our international partners, begin construction on Cosmic Explorer South. The cost is estimated at roughly \$1B USD. The costs for planning, design, and construction can be reduced by leveraging previous investment in Stage 1. We envision using the OzHF facility as the vertex station for Cosmic Explorer South, which may further reduce the cost. All the same, international investment will be essential. We anticipate significant investment from the Asian region in particular, with additional support from the USA, UK and Europe. The Australian host-nation investment would be on the order of \$300M AUD, about a 10\% share in the cost of constructing the entire three-detector, global, third-generation network. After construction has begun on Cosmic Explorer South, the US team upgrades Cosmic Explorer North to Stage 2. Both facilities begin taking data with Stage 2 sensitivity at approximately the same time. Around this time, the Einstein Telescope is taking data in Europe.
 \end{itemize}

\begin{figure*}
    \centering
    \includegraphics[width=1.0\columnwidth]{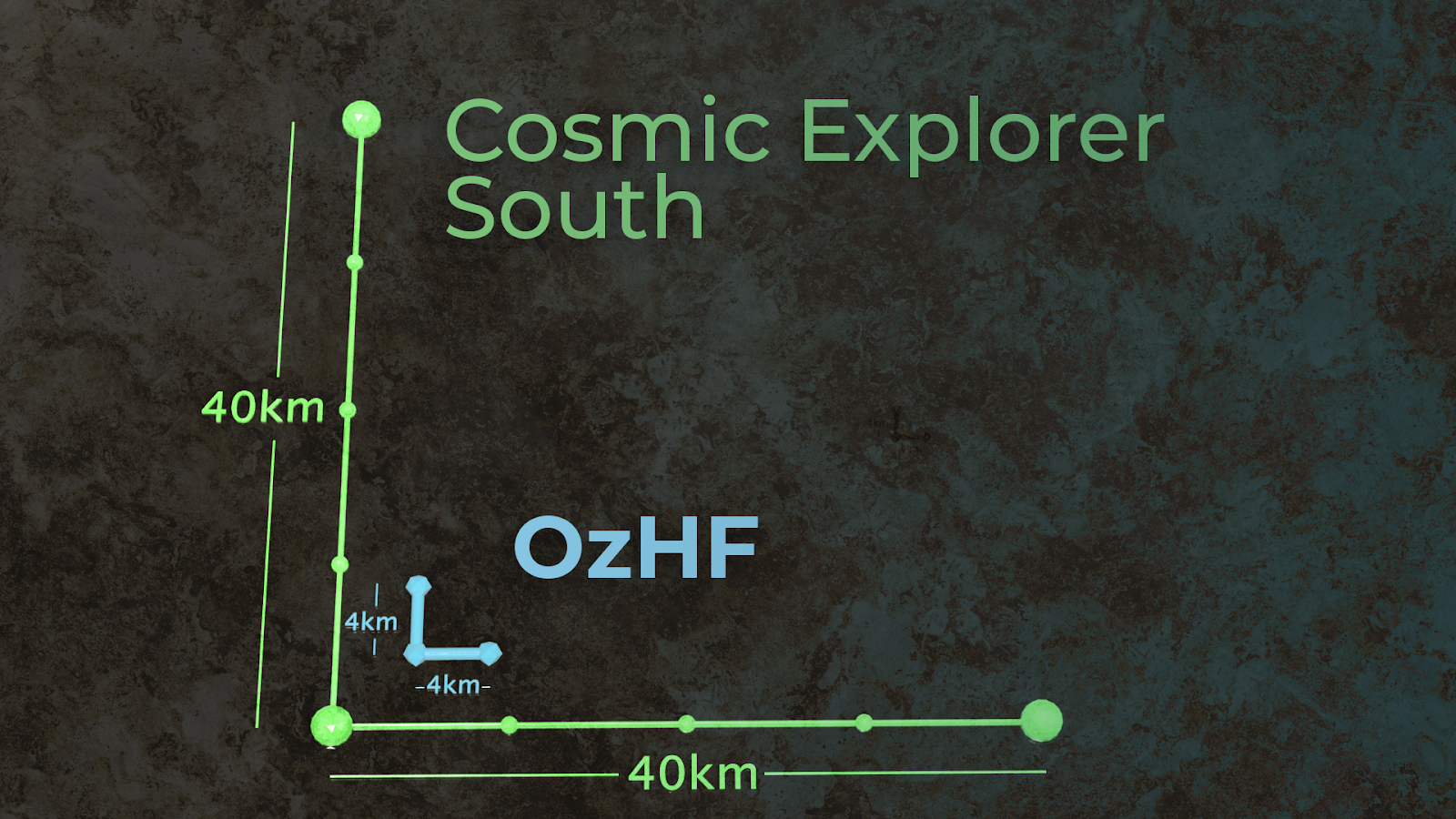}
    \caption{The proposed stages of Australian gravitational-wave detector development. Stage 1 is the ~4 km OzHF observatory, designed to measure the neutron star equation of state while pioneering Stage 2 technology. Stage 2 is Cosmic Explorer South, a gravitational-wave observatory capable of detecting nearly every stellar mass binary merger in the Universe.}
\end{figure*}

\begin{figure*}
    \centering
    \includegraphics[width=0.6\columnwidth]{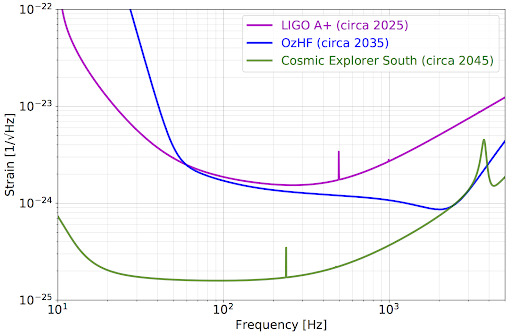}
    \caption{The expected sensitivity (strain noise amplitude spectral density) of OzHF and Cosmic Explorer South.}
\end{figure*}

\begin{figure*}
    \centering
    \includegraphics[width=1.0\columnwidth]{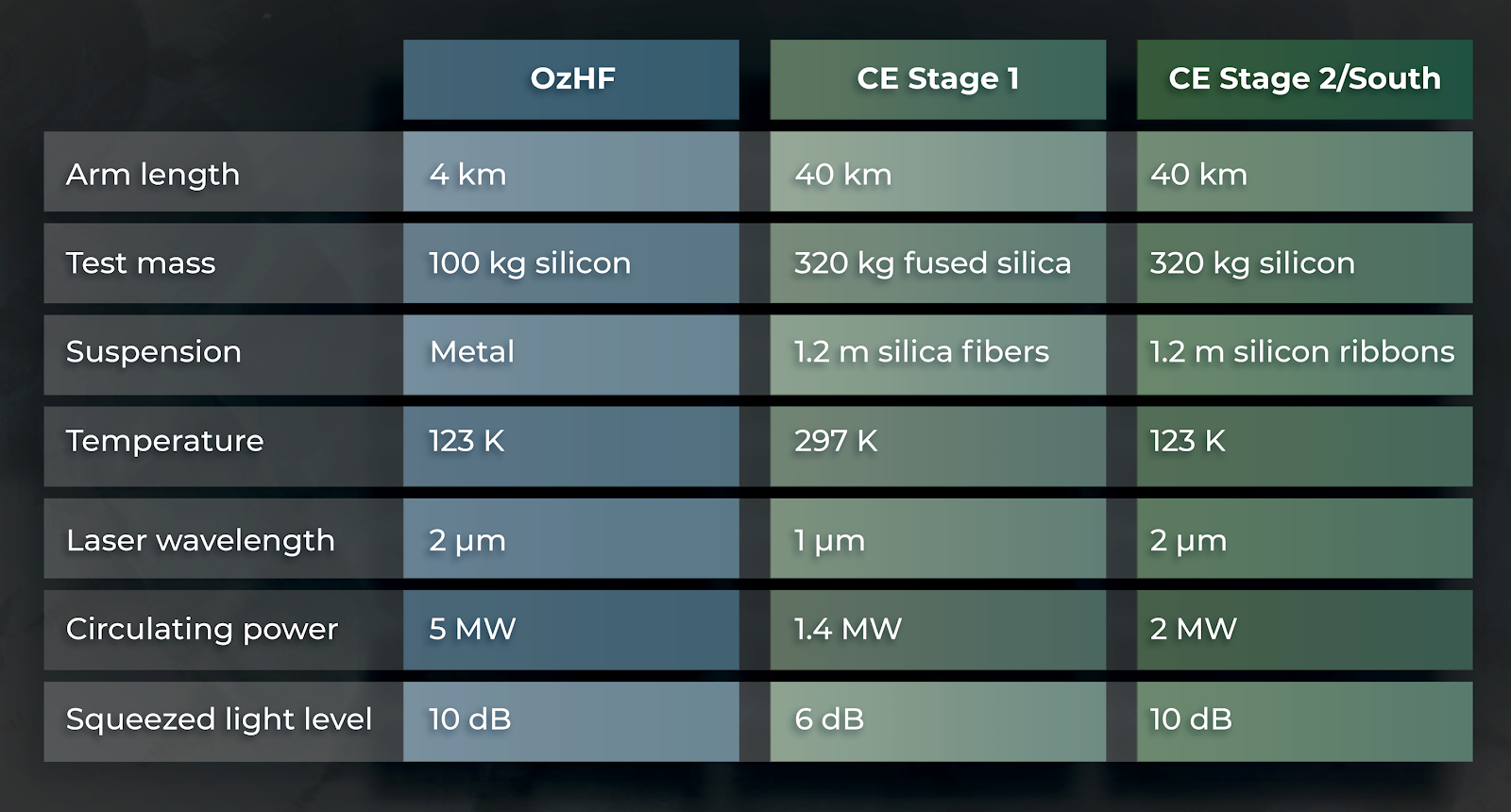}
    \caption{Key technological innovations for the three gravitational-wave observatories discussed in this white paper.}
\end{figure*}

{\bf Part Two: Gravitational-Wave Data Centre.} Astronomy Australia Ltd. (AAL) has provided \$2.8M AUD over two years for the creation of a Gravitational-Wave Data Centre run by Astronomy Data and Computing Services (ADACS). We recommend maintaining this level of investment over the next decade for the ongoing support of Data Centre personnel. This will fund approximately 9 FTE at a cost of \$1.2M AUD/yr as well as \$0.8M AUD for hardware updates roughly every five years. We envisage the data centre working closely with the other astronomy data centres, sharing expertise and personnel time to maximise efficiency.
Current issues and key risks
The most significant risks for the science overview outlined above were retired in 2016 with the announcement that LIGO had detected gravitational waves from a merging binary black hole. This detection confirmed that (1) compact binaries merge frequently, producing gravitational waves detectable with interferometers, and (2) it is possible to design and engineer interferometric gravitational-wave observatories that are limited by a handful of predictable noise sources. The subsequent multi-messenger detection of GW170817 showed conclusively that (3) it is possible to localise electromagnetic counterparts to gravitational-wave detections~\cite{GW170817_mma}. These developments ensure that there is minimal scientific risk for science goals 1-5, which all rely on improved measurements of gravitational waves from compact binaries. Science goal 6, searching for unexpected sources, is subject to the unknown nature of the dark universe.
The Cosmic Explorer facility consists of two, 40 km, ultrahigh-vacuum beam tubes, roughly 1 m in diameter, built in an L-shape on the surface of flat and seismically quiet land. These arms, which are ten times longer than the LIGO arms, will increase the signal-to-noise ratio of gravitational-wave measurements tenfold. In the USA, the plan is to realise Cosmic Explorer observatories in two stages. Cosmic Explorer Stage 1 is expected to use the technology developed for A+, scaled up to a 40 km detector. This provides a relatively low-risk approach to gain at least a factor of ten improvement in gravitational-wave strain sensitivity (and 1000 in event rate).
Cosmic Explorer Stage 2 further improves on Stage 1 with a new set of technologies to reduce the quantum and thermal noises of the detector. Stringent requirements are imposed on materials, laser power, losses, operating temperature, and wavelength. Australian and US R\&D, along with the Australian Pathfinder, will retire much of the risk while delivering important astrophysical results prior to implementation in Cosmic Explorer facilities in the early 2040s.
In addition to these technical risks, there are financial risks. International partnership is essential for Australia to build and operate Cosmic Explorer South. Without investment from international partners, Australia could still provide key technological systems for the Cosmic Explorer network, even if it became untenable to host one of the Cosmic Explorer observatories. Australia would still benefit from being a Cosmic Explorer stakeholder and developer of technology.

\section*{Approximate costings}
A summary of approximate costs is provided in Figures 6 (instrumentation) and 7 (data).

\begin{figure*}
    \centering
    \includegraphics[width=1.0\columnwidth]{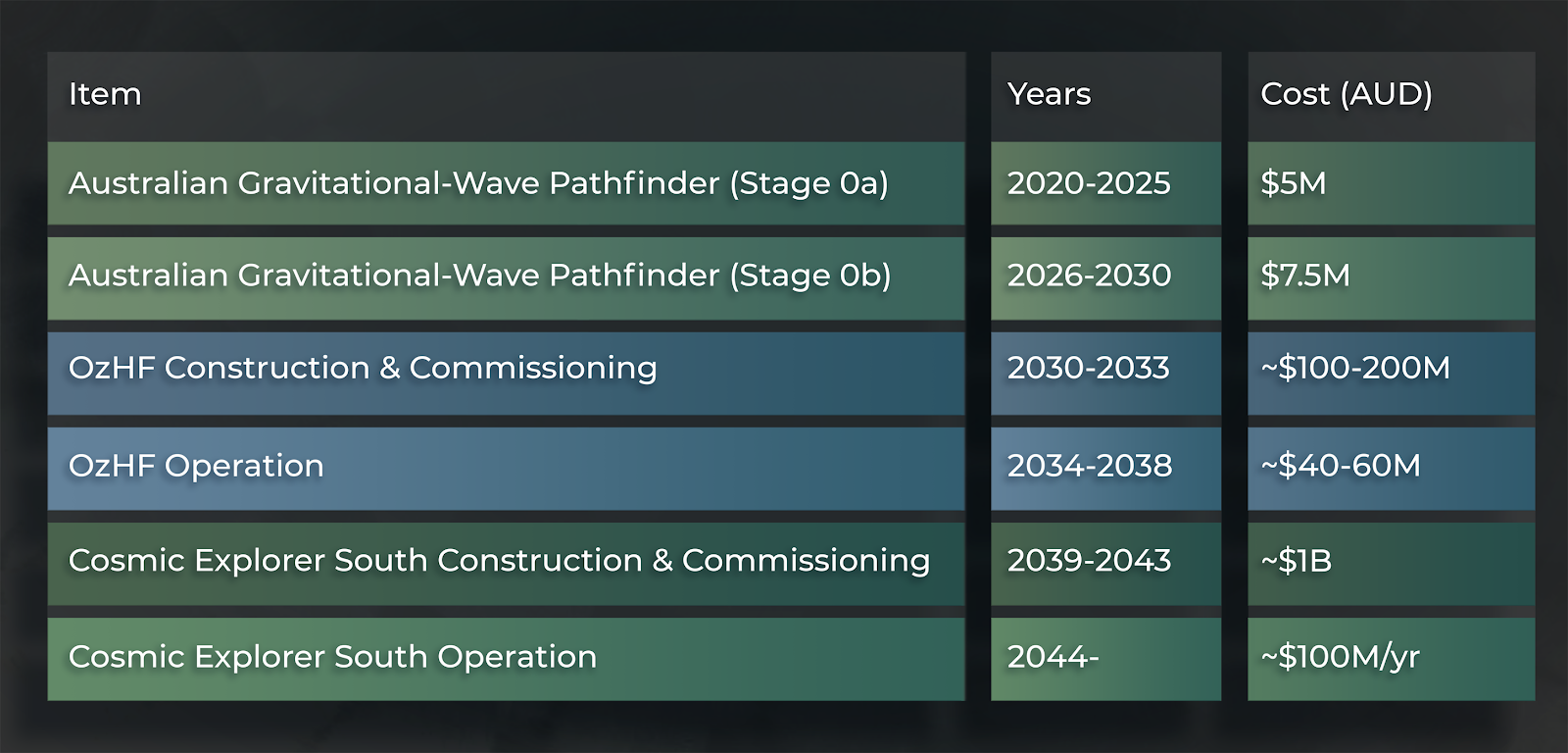}
    \caption{Projected budget for instrumentation including international contributions. The projected budget for the Australian Gravitational-Wave Pathfinder (Stage 0) includes a 10\% contingency. Subsequent costs are highly uncertain. The Pathfinder Stage 0 will produce cost estimates for OzHF and Cosmic Explorer South that are known to within a 10\% contingency.}
\end{figure*}

\begin{figure*}
    \centering
    \includegraphics[width=1.0\columnwidth]{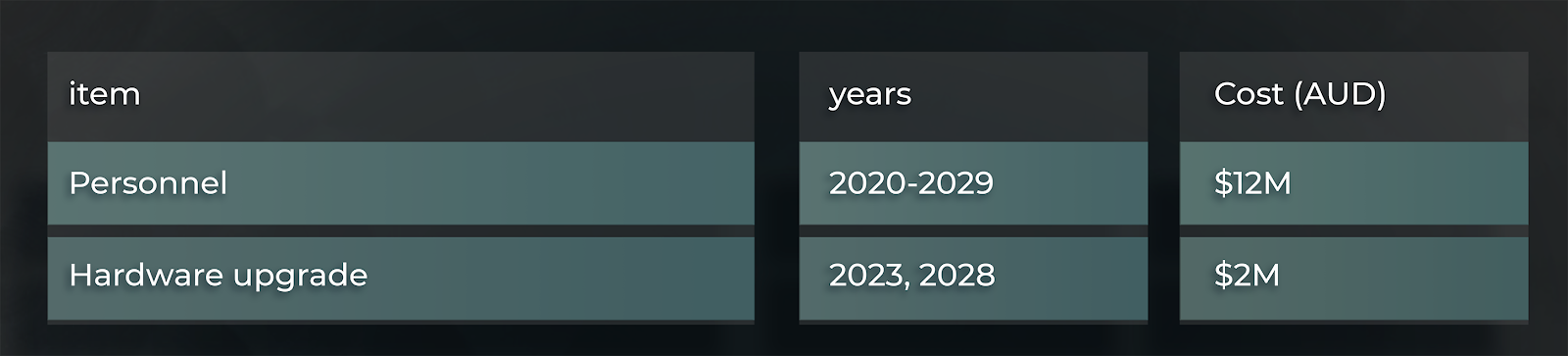}
    \caption{Projected budget for the Gravitational-Wave Data Centre. All entries include a 10\% contingency. The hardware budget is hard-capped.
}
\end{figure*}

\section*{Conclusions \& Recommendations}
It is an exciting time for gravitational-wave astronomy in Australia. The mid-term review should consider investment in (1) a scoping study for the Australian Gravitational-Wave Pathfinder and (2) the Gravitational Wave Data Centre. These investments will ensure continued Australian leadership, developing the technology that enables gravitational-wave astronomers to peer to the edge of the Universe, and interpreting gravitational-wave data to understand the mysteries of the cosmos.

\bibliography{white}

\end{document}